%% file: main.tex
\documentclass[letterpaper, 10pt, conference]{ieeeconf}
\IEEEoverridecommandlockouts
\usepackage{graphicx}
\usepackage{amsmath}
\usepackage{amssymb}
\usepackage{xcolor}
\usepackage{cite}

\title{\LARGE \bf
Internal Feedback in Biological Control: \\ Diversity, Delays, and Standard Theory
}

\author{Josefin Stenberg, Jing Shuang (Lisa) Li, Anish A. Sarma, John C. Doyle%
	\thanks{J. Stenberg is with Engineering Physics, KTH Royal Institute of Technology. J. S. Li and J. C. Doyle are with Computing and Mathematical Sciences, California Institute of Technology. A. A. Sarma is with Computation and Neural Systems, California Institute of Technology. {\tt\small jossten@kth.se, \{jsli, aasarma, doyle\}@caltech.edu}}
	\thanks{This paper is one of three in a series on internal feedback in biological control architectures. These papers may be read in any order, though a suggested order is \cite{Paper1}, then this, then \cite{Paper3}.}
}

\begin{document}

\maketitle
\thispagestyle{empty} 
\pagestyle{empty}

\input{tex_files/01_abstract}
\input{tex_files/02_introduction}

\input{tex_files/04_methodology}
\input{tex_files/05_dess}
\input{tex_files/06_ifp}
\input{tex_files/07_conclusions}

\bibliography{refs/refs_main, refs/refs_muri_doyle, refs/refs_muri_neuro, refs/refs_muri_sls}
\bibliographystyle{IEEEtran}
	
\end{document}


\maketitle
\pagestyle{plain}
\thispagestyle{plain}

\begin{abstract}

Diversity within sensors is visible in many systems. One example is the sensorimotor control system that contains many types of layers that sense with different speed and accuracy. Each component design faces the trade-off between speed and accuracy and the diversity of the system then creates sweet spots with optimal performance. The focus of this project was to develop a model with diverse sensing to explore sweet spots and trade-offs. The aim was to provide examples of diverse systems using LQR and display the difference between only using one certain type of component in comparison with combining components with different properties. For this study two different types of sensors where considered, one type that was fast but sparse and one that was slow but dense. Distinct sweet spots were achieved that presented cases where the different types failed individually and resulted in excellent performance when combined. The obtained solutions contained a lot of internal feedback which proved necessary to avoid instability. It was concluded that diversity is essential for optimal control which agrees well with the expectations. The results of this paper function as a bridge between familiar control theory and the more novel System Level Synthesis framework.

\end{abstract}
\section{Introduction}
In nature and technology there are endless systems and processes we wish to control. Ideally, we want all the features of the sensors to be perfect but that is often infeasible to achieve due to fundamental limitations. One trade-off that is present in most systems is the speed and accuracy trade-off (SAT) which states that what we gain in terms of accuracy, we must compensate for by reducing the control speed \cite{DESS}. Therefore, in a system we often end up with multiple layers of components that are differently fast and accurate. For example a fast and unconscious reflex layer and a slow and conscious planning layer as described in \cite{nakahira2021diversityenabled}.

The combined architecture of these diverse layers creates a final complex system that behaves both fast and accurate which ends up in a so called Diversity Enabled Sweet Spot (DESS) \cite{DESS}. The DESS is a sweet spot caused by the diversity of the components which behaves optimally while all layers still obey the SAT law.

Recent development of the System Level Synthesis (SLS) framework enables the ability to model complicated systems where we can introduce constraints in time and space on the sensing, actuation and communication within the system. It is possible to create and combine diversity within the components which results in intricate systems and the framework design accomplishes this without much computational time due to the scalability of the theory \cite{anderson2019level}. 

There is a lot left to be discovered about diverse sensing before introducing the concepts in SLS. This paper aims to provide further understanding of the sweet spots in diverse systems and present examples of how they can be modeled using senior preSLS control theory frameworks. The focus is to explore the impact of diversity for the fundamental features of DESS using preSLS control theory, as well as to create a bridge between familiar theory and the more novel, powerful features of SLS.

Previous work of Nakahira et al. \cite{nakahira2021diversityenabled}, shows how the sensorimotor control system can be divided into two separate subsystems with a mountain bike example. There is one fast but inaccurate bump-reflex layer and one slow and more accurate trail-planning layer. This example used a scalar formulation 
where the errors of the system were simply added. This formulation abstracted the architectural feature of internal feedback pathways (IFP) and only looked at behavioural DESS. IFP refers to internal feedback loops which are an architectural feature in the system where the components utilizes information about previous actions to predict future actions. The work shown in this paper is different and uses a formulation with multiple states in a diverse system where IFP becomes necessary and the aim is to connect the presence of IFP with DESS.  

A brief introduction of control theory concepts and examples of systems with multiple sensors will be presented in Section \ref{sec:Background}. Section \ref{sec:Methodology} describes the system setup and the resulting model. Analysis of the results will be presented in Section \ref{sec:Results_and_Analysis} and the extension to the SLS framework is described in Section \ref{sec:SLS}. Final conclusions are discussed in Section \ref{sec:Discussions_and_Conclusions}.

\section{Background} \label{sec:Background}
Control theory provides many helpful tools to model the dynamics of a system. For this study a linear–quadratic regulator (LQR) problem has been used to provide a model of a system with diverse sensing and actuation. Some necessary background theory is presented here together with examples of diverse sensing and actuation in reality which motivates the interests of this study.

\subsection{Diverse Sensing and Actuation} \label{sec:DiverseActuation}
The initial example that sparked the desire to explore diverse sensors and actuators is the observation of delays in the sensorimotor control system. Following the example of \cite{nakahira2021diversityenabled}, the sensorimotor control system can roughly be divided into a fast layer with proprioception, and then another slow layer with the visual cortex. The proprioception is our ability to sense the position and movements of our body where the reflexes have a reaction time of around $95.6 \pm 10.6$ ms according to experiments by \cite{proprio}.

The visual reaction time is much slower than the proprioception at around $247.6 \pm 18.54$ ms according to \cite{Jain2015ACS}. Thus suggesting that the part of the sensorimotor control system related to vision has larger delays. The reason is due to the large quantity of computations needed to interpret the information from the raw retinal image. The visual cortex needs to translate a lot of raw data into a decision of which actions needs to be taken \cite{huff_neuroanatomy_2021}.

Although the reaction time is faster, the proprioception is however less accurate than vision and when performing sensorimotor tasks we rely much more on vision for precision and accuracy in the movement \cite{Sarlegna2009}. Thus there exists a SAT obeying system containing diverse components where one layer is fast and inaccurate which together with a slow and accurate layer creates a DESS. 

Within vision there is a lot of internal feedback between the components of the visual network and IFP is commonly found in the visual area \cite{Felleman91distributedhierarchical, CALLAWAY2004625, MUCKLI2013195}. In general the world is slow changing in relation to the $\sim 200$ ms it takes for us to see everything in a room. Using IFP, our vision can use predictions to compute error signals which are smaller and can be sent faster than the full raw image. The internal delays of vision also result in stale images by the time they arrive to the motor areas which also motivates the benefits of the predictions from IFP. There are also experiments which point to IFP from motor to visual areas \cite{Churchland}.

Another motivating example of diverse sensing and actuation is the immune system response. The complexity of the immune system is undeniable. The components of the immune system protect us from pathogens and can initially be divided into two parts: the innate immunity and the adaptive immunity \cite{immunology}. The innate immunity provides a fast but not very accurate protection and is able to limit infections to some extent. Here cells like macrophages, neutrophils and basophils respond within minutes to hours with a rapid response. The adaptive immunity is the accurate but slow defense with specialized protection against pathogens. In comparison the adaptive immunity may take weeks to respond but when it does, it works very effectively with T lymphocytes and B lymphocytes that produces an efficient and long lasting protection \cite{immunology}. Thus the immune system also presents a DESS system with a fast but inaccurate layer and an accurate but slow layer.

IFP is also visible in the immune system where T cells can be further divided into CD4+ T cells, or T helper cells, and CD8+ T cells, or cytotoxic T cells, which function as the names suggest. The CD8+ T cells kills abnormal or damaged cells, which can be virus infected or cancerous \cite{CD8cancer}. The CD4+ T cells on the other hand help regulate the responsiveness of the killer cells and promotes interaction in between cells in the immune system \cite{CD4}. The T cells are activated via negative feedback \cite{immunefeedback} suggesting a lot of IFP present in the immune system in order to recognize and be sensitive to changes and irregularities.

The ubiquity of SAT, DESS and IFP in both the sensorimotor system and the immune system motivates us to seek fundamental theory on the optimal usage of diverse components (e.g. diverse sensors and actuators). The current goal is not to model these biological systems in fine detail, but to study simple, human-interpretable examples that give rise to fundamental theory on the existence and characteristics of DESS and IFP.

In reality most of the delays are caused by limitations in communication between components. For this study the origins of the delays are simplified. The idea is to push all delays into the sensors and actuators to fit the preSLS theory. Given the abstractions and simplifications the goal is to investigate if it is possible to find stories that shows DESS and IFP using preSLS theory in a simple and approachable way. 

\subsection{LQR} \label{sec:LQR} 

Linear-quadratic optimization refers to a problem formulation where the objective is to minimize a cost that is described by a quadratic function, as in Equation \eqref{eq:cost}, and the system dynamics are described by linear differential equations, as in Equation \eqref{eq:dynamics} \cite{GladTorkel2000Ct:m}. Here $x_k$ denotes the state vector, $u_k$ the input, $y_k$ the output and the disturbance $w_k$ is assumed to be Gaussian distributed. 

\begin{equation} \label{eq:cost}
\begin{aligned}
    J = \lim_{\tau \to \infty}\sum_{k=1}^{\tau} x_k^T Q x_k + u_k^T R u_k
\end{aligned}
\end{equation}

\begin{equation} \label{eq:dynamics}
\begin{aligned}
x_{k+1} & = A x_k + B_1 u_k + B_2 w_k \\
y_k & = C x_k  
\end{aligned}
\end{equation}

The optimal control is given by $u_k=-K x_k$ where $K$ is the optimal controller and is given by: $$K=(R + B_1^T P B_1)^{-1}(B_1^T P A)$$ and $P$ is solved for using the discrete time algebraic Riccati equation (DARE):

\begin{equation}
\centering
\begin{aligned}
P &= A^T P A - (A^T P B_1)(R + B_1^T P B_1)^{-1} (B_1^T P A) + Q
\end{aligned}
\end{equation}

For this study only the dual problems of State Feedback (SF) and Full Control (FC) will be considered where either $C=I$ or $B_1=I$ respectively in Equation \eqref{eq:dynamics}. The general Output Feedback (OF) case will not be considered and is left for future studies. Additionally, the sensor noise is assumed to to be zero which is equivalent to $R=0$ in Equation \eqref{eq:cost}.

One key property of the states is the concept of controllability. A state $x$ is said to be \textit{controllable} if in finite time, an input can yield the desired state $x$. If all states are controllable then the whole system is controllable \cite{GladTorkel2000Ct:m}.

A related theorem states that a system of order $n$ is controllable if and only if the controllability matrix $\mathcal{S}$ has full rank where the controllability matrix is given by: $$\mathcal{S}(A,B)= \begin{bmatrix}
B & AB & A^2B & \dots & A^{n-1}B
\end{bmatrix}$$ Here $A$ and $B$ are the matrices of Equation \eqref{eq:dynamics} \cite{GladTorkel2000Ct:m}. 

The properties of this theorem yields a trivial case where actuation density is demanded. If $A=I$ then the stability matrix becomes: $$\mathcal{S}= \begin{bmatrix}
B & B & \dots & B
\end{bmatrix}$$ Thus $B$ must be of the same rank as $A$ for the system to be controllable, meaning that for each state we must also have an actuator. This density requirement of the actuators is considered trivial and will not be further discussed. The aim is instead to look at diversity both in time and sparseness to see how diversity can yield stable control. 

\section{Methodology} \label{sec:Methodology}
We assume the system setup to be a ring formation with n states as shown in Figure \ref{fig:ringformation}. All interactions between neighboring states and own interactions with previous time steps are assumed to be equal to some constant which yields the $(n \times n)$ $A_1$ matrix describing the dynamics:

\begin{equation}
    A_1 = \frac{a}{3} *
    \begin{bmatrix} 
    1 & 1 & 0 & \dots & 1\\
    1 & 1 &  1 & 0 &  \dots  \\
    0  & \ddots & \ddots &  \ddots &  \vdots \\
    \vdots & \ddots & \ddots & \ddots & 1\\
    1 &     0   & \dots & 1 &  1
    \end{bmatrix}
\end{equation}
$a$ is a scaling parameter which scales the eigenvalues of the $A_1$ matrix and has been chosen to make $A_1$ neutrally stable for $a=1$, i.e. the maximum eigenvalue is equal to 1. The scaling parameter $a$ can be adjusted to make eigenvalues of the $A_1$ matrix become $>1$ which creates an unstable system. 

\begin{figure}
\centering 
\includegraphics[width=0.4\linewidth]{plots/1_ringform.png}
\caption{The ring formation with $n$ states where each state is connected to two neighbors.}
\label{fig:ringformation}
\end{figure}
We consider two types of sensors: one that is fast but sparse, and one that is slow but dense, motivated by the examples discussed in Section \ref{sec:Background}. Sparse and dense in this context refers to the quantity of sensors or actuators in the system in comparison to the states. Dense actuation means that we have as many actuators as states which can act on all states at once. Sparse actuation on the other hand means that the control is limited to a smaller section of the system.

\subsection{Diverse Sensing}
Suppose that we have a time discrete system with the following dynamics: 

\begin{equation} \label{eq:FC}
\begin{aligned}
x_{k+1} & = A x_k + B_2 u_k + B_1 w_k \\
y_k & = C x_k
\end{aligned}
\end{equation}
where $B_2 = I$ for the diverse sensing FC case. The first $n$ entries of $x_k$ are the actual, real states and the rest denote the delayed, internal sensed states. Here $B_1 = \begin{bmatrix}
I & 0 & \dots
\end{bmatrix}$, where $I$ is the identity with dimension $n$ and $d$ blocks of zero.

The sensed output is divided into two parts, one that is fast and sparse, and another which is slow and dense. The corresponding parts of the $C$ matrix, $C = [C_f \quad C_s]^T$, are denoted $C_f$ for fast and $C_s$ for slow. The format of the $C$ matrix depends on two additional parameters besides the number of states $n$: the relative delay $d$ between the fast and slow sensing, and the dimension $m$ of the fast sensing, which is how many $m$ eigenvectors of $A_1$ that are included in the fast component.

$C_f$ is of the size $(n\cdot (d+1) \times m)$ where the top $(n \times m)$ block is a matrix with the $m$ eigenvectors corresponding to the $m$ largest eigenvalues of $A_1$. The remaining entries of $C_f$ are zero: 

\begin{equation}
    C_f =
    \begin{bmatrix} 
    E & 0  & \dots & 0  \\
    \end{bmatrix}^T 
\end{equation}
where we let E be the block matrix of the $m$ eigenvectors of $A_1$. Thus $C_f$ represents the instant but sparse sensing which senses in the $m$ "worst directions" of the spread when given an impulse to the system. 

$C_s$ is of the size $(n\cdot (d+1) \times n)$ which contains $d$ stacked $n \times n$ blocks of zeros and the bottom $n\times n$ block is the identity. 

\begin{equation}
    C_s =
    \begin{bmatrix} 
    0 & 0  & \dots & I  \\
    \end{bmatrix}^T 
\end{equation}

Thus $C_s$ represents dense sensing that is delayed $d$ time steps. To include the delayed states of the system, the extended matrix $A$ then becomes:

\begin{equation}
    A =
    \begin{bmatrix} 
    A_1 & 0 & \dots & 0  \\
    I & 0 &  & \vdots \\
    0 & \ddots & & \\
    \vdots & \dots  & I & 0
    \end{bmatrix}
\end{equation}
where a delay $d$ yields the dimension of $A$ as $(n\cdot (d+1) \times n\cdot (d+1))$.

We now formulate an LQR problem with $Q=G^T G$ and $R= 0$ where $G^T = B_2$. Thus only the $n$ original states are minimized in the objective function and the delayed internal states are excluded. 

The optimal control action is $u_k=-L y_k = -LC x_k$, where $L= A^T F C ( C^T F C + R )^{-1}$ is obtained by solving DARE.

The full control problem described in Equation \eqref{eq:FC} and the LQR weights $G^T G$ and $R=0$ will be referred to as the \textit{(FC) problem}. With no noise the cost of the objective is calculated as $\text{Tr}(B_2^T F B_2)$.

The case of diverse actuation is dual to that of diverse sensing. Due to the duality of the two problem formulations the cost is equal for both cases. 

Later we study the behaviour of the systems with diverse sensing when exposed to an impulse in the first time step. The aim is to search for trade-offs and DESS by varying the different parameters and to find interesting examples showing variations of the performance when using the fast and slow components separately and combined.

\section{Results and Analysis} \label{sec:Results_and_Analysis}
The system setup has four parameters which affect the outcome: the number of states $n$, the scaling parameter $a$, the delay of the slow component $d$ and the sparseness of the fast component, which is determined by how many $m$ eigenvectors of $A_1$ are included. The open loop with no control results in an unstable system for $a\geq 1$ since for these values the eigenvalues of $A_1$ become larger than or equal to 1. For these occasions an impulse given at any place in the ring will cause the states to grow rapidly. 

\subsection{Diverse Sensing}
As described above, we consider two types of sensors: fast and slow. First, we study system performance in the absence of diverse sensors. We show that both fast-only and slow-only sensing, on their own, can perform terribly and incur extreme (sometimes infinite) costs. We then show that by allowing the system to use diverse sensors (i.e. both fast and slow), performance is dramatically improved. This demonstrates the existence of a striking DESS.

\subsubsection{Fast-only sensing}
For $m=1$, the relationship between the objective value and the value of $a$ is shown in Figure \ref{fig:fast_obj_vary_a_m1}. 
\begin{figure}
\centering
\begin{subfigure}{0.49\linewidth}
\centering
\includegraphics[width=0.99\linewidth]{plots/2a_fast_obj_vary_a_m1.png}
\caption{}
\label{fig:fast_obj_vary_a_m1}
\end{subfigure}
\begin{subfigure}{0.49\linewidth}
\centering
\includegraphics[width=0.99\linewidth]{plots/2b_a_vs_n.png}
\caption{}
\label{fig:4b_a_vs_n}
\end{subfigure}
\caption{Relations of $a$ and $n$ where fast-only results in infinite costs for $m=1$: (a) Objective value as a function of $a$ for different $n$. (b) $a$ values where fast-only fails as a function of $n$.}
\label{fig:avals}
\end{figure}
It is clear how the function of the objective value becomes almost vertical for certain values of $a$ which shows the breaking points of where the fast-only sensing fails and results in infinite cost. Here more than $m$ eigenvalues become larger than or equal to 1 and for which the eigenvectors in the fast-only sensing are no longer sufficient to cover all the impulse propagation directions.

Figure \ref{fig:4b_a_vs_n} shows the breaking point values of $a$ for which the fast-only sensing with $m=1$ results in infinite costs. Here it is evident that as the $n$ number of states increase, the system is less able to control instability.

If we examine Figure \ref{fig:fast_obj_vary_a_m1} we get that for $n=5$, $m=1$ the fast-only sensing fails for $a=1.856$. This also means that the fast-only actuation fails for any $a\geq1.856$ too. Thus this breaking point shows the minimum instability measure where fast-only starts to fail. The corresponding impulse response for this setup using fast-only sensing is show in Figure \ref{fig:fastact_a1856_n5} where we note how the impulse response does not get attenuated to zero.

The fast-only component can only sense in the "worst case" direction and for this setup the second and third eigenvalue has become 1 which yields this step response where the states become constant. Hence, the fast-only component manages to attenuate the exponentially growing part of the system but can not act on the constant steps which yields the infinite cost. 

\subsubsection{Slow-only sensing}
We now consider the case of slow-only sensors which are delayed $d=3$ time steps. If the setup with $n=5$, $m=1$ $a=1.856$ is repeated when only allowing slow actuation with $d=3$, the system behaves as shown in Figure \ref{fig:slowsense_n5_a1856_m1}.

\begin{figure}
\begin{center}
\begin{subfigure}{0.31\linewidth}
\centering
\includegraphics[height=0.7\linewidth]{plots/3a_fastact_a1856_n5.png}
\caption{Fast-only}
\label{fig:fastact_a1856_n5}
\end{subfigure}
\begin{subfigure}{0.31\linewidth}
\centering
\includegraphics[height=1.5\linewidth]{plots/3b_slowsense_n5_a1856.png}
\caption{Slow-only}
\label{fig:slowsense_n5_a1856_m1}
\end{subfigure}
\begin{subfigure}{0.31\linewidth}
\centering
\includegraphics[height=1.5\linewidth]{plots/3c_fastandslowsense_n5_a1856_m1.png}
\caption{Fast \& slow}
\label{fig:fastandslowsense_n5_a1856_m1}
\end{subfigure}
\caption{Impulse responses for $n=5$, $a=1.856$, $m=1$ and $d=3$ with cost = a) $\infty$ b) 13.726, c) 2.279.}
\label{fig:sense_n5_a1856_m1}
\end{center}
\end{figure}

Note that the top 5 states of Figure \ref{fig:slowsense_n5_a1856_m1} display the actual states where as the bottom states are the internal, delayed states. For all following impulse responses, the top 5 states are the actual ring states of the system and are encircled with green for clarification. The dashed lines display the distinction between the delayed states. 

It is clear that the first three time steps behaves like open loop, and the slow-only actuation behaves like there is not any control action present. This is due to the delay of the slow sensors. Once they are caught up, the impulse propagation is killed off immediately, showing the effects of the slow sensor being dense. 

\subsubsection{Fast \& Slow sensing}
The combination of the fast and slow senors is what is of most interest since this setup is what creates the diversity of the system. The impulse response when utilizing both types of actuators is shown in Figure \ref{fig:fastandslowsense_n5_a1856_m1}. 

This case shows a combination of the behaviors from both previous setups where the initial impulse response in the first three time steps are equal to the fast-only response and then all impulse propagation is attenuated completely when the slow sensor has responded. 

Since the fast-only sensing yields infinite cost for $a=1.856$ we fix the scaling parameter $a$ as this breaking point and study the objective value of the slow-only compared with the fast \& slow sensing while varying the delay $d$. The corresponding result is shown in Figure \ref{fig:n5_m1_vary_d}.
\begin{figure}
\centering
\includegraphics[width=0.7\linewidth]{plots/4_n5_m1_vary_d_obj_val.png}
\caption{Objective value of fast-only, slow-only and fast \& slow sensing for $n=5$, $a=1.856$ and $m=1$.}
\label{fig:n5_m1_vary_d}
\end{figure}

\subsubsection{DESS}
A dramatic sweet spot has been found. The results are clearly showing how much better the performance is when combining the two different sensors in comparison with using only fast or slow sensing separately. This striking sweet spot shows how the mixture of fast and slow sensing results in a low cost at the same time as the fast-only sensing results in infinite costs and the costs of the slow-only sensing is increasing exponentially when the delay increases.

 Another really important result is the difference in the worst case maximum amplitude for the different cases. The slow-only case yields in an infinite maximum amplitude as the delay increases due to that the slow-only impulse acts as the open loop until the delayed slow sensing may operate. When using the combination of fast \& slow, the maximum amplitude instead remains to be close to the impulse value. Thus, we can conclude that delayed sensing is acceptable if combined with a fast but sparse sensor. Hence, the theoretical sweet spot found here is consistent with what real examples, such as the sensorimotor control system, suggest.
 
 Similar sweet spots are obtained for higher $n$ and smaller $a$ following the scheme of Figure \ref{fig:fast_obj_vary_a_m1}. However, the decrease of the $a$ value is generating a less dramatic sweet spot where the delay $d$ needs to increase a lot in order for the fast \& slow sensing to greatly outperform the slow-only sensing. Decreasing the scaling parameter $a$ is generating a system with slower dynamics which in turn requires the delay to increase in order to demonstrate a visible effect. 
 
The fast-only sensing becomes more sensitive to instability as $n$ increases. When increasing $m$, more eigenvectors are included corresponding to the largest eigenvalues in descending order. In fact, the fast-only sensing can achieve a stable system as long as $m$ is larger than or equal to the number of eigenvalues of $A_1$ that are $\geq1$. Thus, intuitively, decreasing the sparseness of the fast sensing improves the performance in terms of stability. With larger $n$ and $m$ similar sweet spots are observed where the diversity enables much better performance.

We previously considered the case of perfect actuation and diverse sensing. Mathematically, this is the dual problem to that of perfect sensing and diverse actuation; thus, our observations about diverse sensing and DESS also apply to diverse actuation. 

This result is displaying dramatic sweet spots where the combination of both fast and slow sensing is essential for the system to function properly. It is showing that delays in actuation are acceptable if they are combined with a fast and sparse response. Even in a very simple system with only two types of actuators, dramatic DESS are observed, suggesting that DESS is a fundamental feature in the presence of diverse sensing. 

\subsection{IFP}
We now observe the structure of the controllers that give the dramatic DESS found above. We show that IFP is an essential architectural feature to enable DESS and that the removal of it worsens the performance of the system.

Looking at the impulse responses of Figure \ref{fig:slowsense_n5_a1856_m1} and Figure \ref{fig:fastandslowsense_n5_a1856_m1} there seems to be a lot of IFP present, meaning that there is information flow traveling backwards to the sensors within the system. This is displayed clearly in Figure \ref{fig:Kpics} which shows the optimal controllers $L$ of the diverse sensing problem. The second, third and fourth blocks of $L$ show the IFP of the system. Here the entries of the matrix is shown using a color map to more easily visualize patterns rather than showing large matrices full of numbers.
\begin{figure}
    \begin{center}
    \begin{subfigure}{0.49\linewidth}
    \includegraphics[height=\linewidth]{plots/5a_KT_a1856_m1_n5_slow.png}
    \caption{Slow-only}
    \label{fig:Kslow}
    \end{subfigure}
    \begin{subfigure}{0.49\linewidth}
    \includegraphics[height=\linewidth]{plots/5b_KT_a1856_m1_n5_fastandslow.png}
    \caption{Fast \& slow}
    \label{fig:Kfast}
    \end{subfigure}
    \end{center}
\caption{The optimal controllers $L$ for the diverse sensing setup with $n=5$, $a=1.856$, $m=1$ and $d=3$}
\label{fig:Kpics}
\end{figure}

Figure \ref{fig:Kslow} displays the IFP of the case where only the slow but dense sensing is used. The shape of the $A_1$ matrix is clearly visible, which is not a coincidence since this is what the internal feedback can utilize in terms of predicting the behaviour of the states in the system. Comparing with the case where both fast and slow sensors are used in Figure \ref{fig:Kfast}, the shape is still similar although including the fast sensor changes the magnitudes of the entries in the controller corresponding to the slow and dense sensor. 

The magnitude of the entires in $L$ in Figure \ref{fig:Kslow} are much grater than in Figure \ref{fig:Kfast} suggesting that there is much more IFP in the slow-only case. Introducing the fast component in Figure \ref{fig:Kpics} greatly reduces the need for IFP in the delayed states since the fast action is taken and this seems to dominate over the now smaller slow-only based action. It is also clear that there is less IFP present the more delayed the internal state is, i.e. the magnitudes in the blocks of $L$ are decreasing for each delayed step. However, for both cases it is clear that the system uses IFP to create optimal control and it is necessary with internal feedback to know which action has been taken previously such that the following delayed states does not act poorly.

\subsubsection{Removing IFP}
To further study the impact of IFP we study the case where it is removed in the controllers $L$ for both the slow-only and the fast \& slow setup. This is done by removing the corresponding IFP blocks of $L$ and setting the to be zero.

The removal of IFP causes dramatic change. Figure \ref{fig:slownoifp} shows the impulse response when using slow-only sensing with removed IFP where the setup is still $n=5$, $a=1.856$, $m=1$ and $d=3$. 

The impulse causes major fluctuation in the states due to the delayed sensing. Removing IFP causes a gap in the system information about the current state and removes the knowledge of previous actions, which before could be determined perfectly using the internal feedback. Without the IFP, the system is no longer are able to calculate the correct action since the internal feedback of what has previously been done is removed. This generates a system which only acts based on what has been sensed $d$ time steps ago without taking previous, more recent actions into account.

\begin{figure}
    \begin{center}
    \begin{subfigure}{0.49\linewidth}
    \centering
    \includegraphics[height=1.2\linewidth]{plots/6a_no_ifp_slow_a1856_response.png}
    \caption{Slow only sensing}
    \label{fig:slowsense_n5_a1856_no_ifp.png}
    \end{subfigure}
    \begin{subfigure}{0.49\linewidth}
    \centering
    \includegraphics[height=1.2\linewidth]{plots/6b_fastandslowsense_noifp.png}
    \caption{Fast \& slow sensing}
    \label{fig:fastandslowsense_noifp.png}
    \end{subfigure}
    \end{center}
\caption{Slow-only and fast \& slow sensing with removed IFP when given an impulse with $n=5$, $a=1.856$, $m=1$, $d=3$ and cost = a) $\infty$, b) $\infty$.}
\label{fig:slownoifp}
\end{figure}

The behavior is clearly visible in the impulse response where the impulse is attenuated perfectly after $d+1$ time steps when the delayed sensing has caught up with the delay. But immediately in the next time step the states are fed with more attenuating control action which causes the 0 states to become negative. This is due to that the sensed state $d$ time steps ago was still positive and thus negative actuation was needed and without the internal feedback there is no way of telling that this has already been taken care of in the previous time step. 

Thus, more negative actuation is fed into the system causing the states to become negative. Later, these negative states will be sensed and positive action will be taken, but due to the delay and lack of internal feedback the entire system will begin to fluctuate between positive and negative states, each time growing in magnitude. Hence without IFP the system becomes unstable. 

\begin{figure}
\centering
\includegraphics[width=0.6\linewidth]{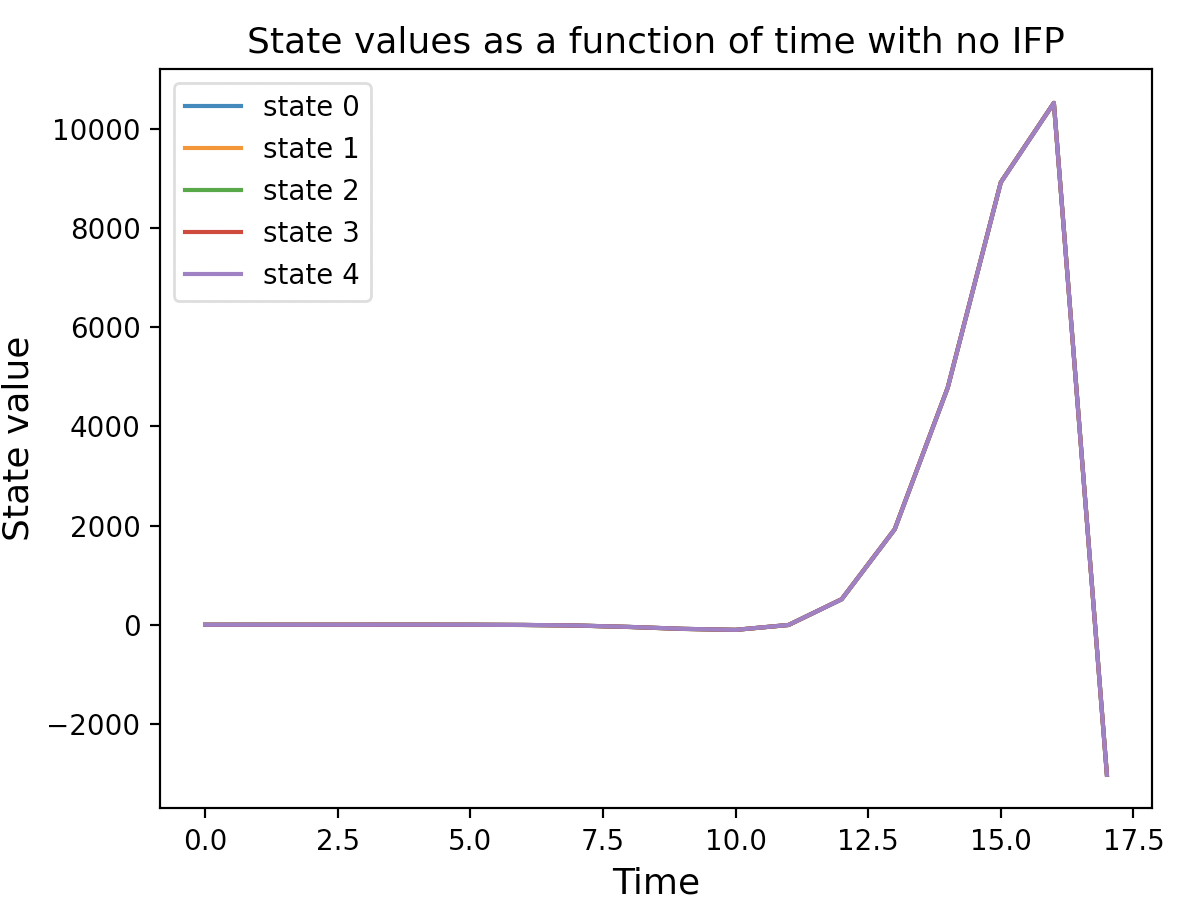}
\caption{Slow-only sensing or actuation with removed IFP when given an impulse.}
\label{fig:slowsense_n5_a1856_no_ifp_states.png}
\end{figure}

This fluctuating behavior for slow-only is shown in Figure \ref{fig:slowsense_n5_a1856_no_ifp_states.png}. Due to the increase in magnitude, the fluctuation appears as a flat line when in fact the states are fluctuating with exponential increase in magnitude. Since the ring states are equal for the dual cases, the figure applies to both slow-only cases of sensing and actuation. 

Removing IFP in the fast \& slow sensing case yields the impulse responses as shown in Figure \ref{fig:fastandslowsense_noifp.png}. The impulse response again fluctuates such that the amplitude increases to infinity.

Comparing the results of slow-only and fast \& slow, it is evident that the system using both kinds of components behaves better since the magnitude of the state values increase less rapidly. However, for long time intervals, the amplitude of the states will increase to infinity suggesting that IFP is a crucial attribute of the system in order to achieve stability. 

What is interesting to note is the result of the fast \& slow without IFP for cases where the fast-only component alone is sufficient for the system to achieve stability. There the fluctuations are decreasing in amplitude and for long time intervals the impulse is attenuated to 0. However, the impulse attenuation for fast-only is much faster. Thus, including the slow components without IFP makes the system performance worse because of the delayed sensing without feedback, suggesting the importance of IFP. 

Removing IFP from either the slow-only or fast \& slow case results in worse performances of the system. The setup that resulted in sweet spots in previous figures, yielded an unstable system with infinite magnitudes without the internal feedback pathways thus highlighting the importance of its presence. 

Without IFP, the slow-only setup will always result in instability because of the delayed reaction. But even with both fast and slow components the system dynamics are more limited and the scaling parameter $a$ needs to be small in order for the system to remain stable. Thus signaling the importance of IFP when delays are present in order to reduce control limits. It is also evident that IFP is a necessary architectural feature to enable DESS in the presence of delayed layers. 

Given the experience of these results that are presented in this paper, it is clear that the preSLS control theory is sufficient to describe a simplified model of diverse systems, and which functions as a bridge between familiar theory and future extensions. The key takeaways are that DESS is a fundamental feature generated by the diversity of the systems and that IFP is a fundamental feature necessary to enable DESS when delays are present.

\section{Extension to SLS} \label{sec:SLS}
Moving forwards there are two directions of further studies. One direction is to extend all of the experiences from the senior, preSLS control theory to the SLS framework. Using the understanding from the fundamental examples, more complex applications to the sensorimotor control system or immune system can be made using SLS. There we still assume full state feedback with full control but the delays are moved from the sensors and actuators. Instead, internal communication and computation delays are imposed which causes an explosion in IFP which demands a switch to the SLS framework. 

The first key result of this research is that large delays in sensors or actuators, which exists in the visual cortex, are acceptable if the components are dense and accurate and if these delayed components are complemented by sparse and fast components, like the vestibular system and proprioception. Extending the implementation of diverse actuation in SLS is easily done by first creating a concatenated $\mathcal{B} = \begin{bmatrix}
B_1 & B_2 & \dots
\end{bmatrix}$ matrix with all the $B:i$ matrices describing the separate dynamics of each type of actuator. Secondly the delayed parts of the diverse actuation is implemented by constraining the corresponding parts of the delayed actuator in each block matrix of the $\Phi_u$ matrix from \cite{anderson2019level}. The constraint of diverse, delayed actuation and sensing can then be further analyzed in combination with implementing locality and communication constraints using the SLS framework where the scalability property is essential to be able to describe further complex systems.

The second key result is regarding the presence of IFP in the optimal controllers for the delayed components. The necessity of IFP for a well-functioning system is a good motivator for continued studies of these kinds of systems using SLS. As the size and complexity grows, the more need for massive IFP, which will affect the computational time. Hence when trying to model systems as the sensorimotor control system or the immune system, that are exhibiting functions with a lot of IFP, the scalability and locality of the SLS theory is crucial. 

The other direction of further studies would be to combine the two separate cases of diverse sensing and actuation by switching to OF dynamics, introduce sensor noise and then use a Kalman filter \cite{GladTorkel2000Ct:m} to estimate the state of the system. This could be done with preSLS theory or with SLS to extend the previously mentioned direction of future studies even further.

\section{Discussions and Conclusions} \label{sec:Discussions_and_Conclusions}
The results of this paper show that there exists dramatic sweet spots in systems with diverse sensing and actuation where the components fail to control the system separately but create a well-functioning setup when combined. The results also show the importance of internal feedback within delayed components where the system performance becomes unstable without the correct feedback information. Examples of applications where the theory of diverse sensing and actuation can be implemented include the sensorimotor control and the immune system. These are all complex structures which when modeled requires large computations. This yields SLS as a natural future direction of study.

\addtolength{\textheight}{-12cm}  

\bibliography{References}

%% file: tex_files/01_abstract.tex
\begin{abstract}

Neural architectures in organisms support efficient and robust control that is beyond the capability of engineered architectures. Unraveling the function of such architectures is challenging; their components are highly diverse and heterogeneous in their morphology, physiology, and biochemistry, and often obey severe speed-accuracy tradeoffs; they also contain many cryptic \textit{internal feedback pathways} (IFPs). We claim that IFPs are crucial architectural features that strategically combine highly diverse components to give rise to optimal performance. We demonstrate this in a case study, and additionally describe how sensing and actuation delays in standard control (state feedback, full control, output feedback) give rise to independent and separable sources of IFPs. Our case study is an LQR problem with two types of sensors, one fast but sparse and one dense but slow. Controllers using only one type of sensor perform poorly, often failing even to stabilize; controllers using both types of sensors perform extremely well, demonstrating a strong \textit{diversity-enabled sweet spot} (DESS). We demonstrate that IFPs are key in enabling this DESS, and additionally that with IFPs removed, controllers with delayed sensing perform poorly. The existence of strong DESS and IFP in this simple example suggest that these are fundamental architectural features in any complex system with diverse components, such as organisms and cyberphysical systems.
\end{abstract}

%% file: tex_files/02_introduction.tex
\section{Introduction}

\begin{figure}
\centering
\includegraphics[width=7cm]{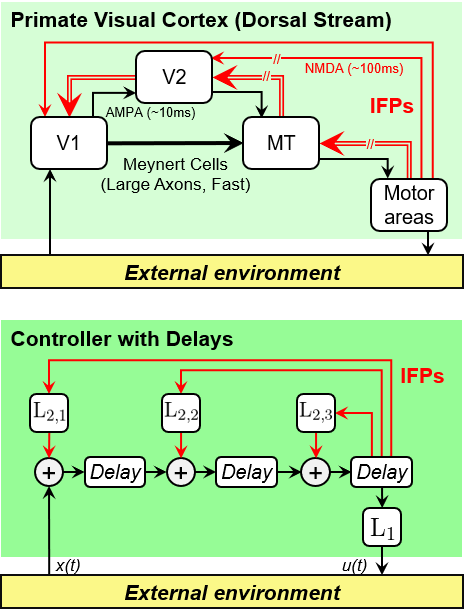}
\caption{Forward pathways (in black) carry information from sensing toward actuation (left to right). Internal feedback pathways (IFPs, in red) travel in the reverse direction. \textit{(Top)} Primate visual cortex contains large amounts of IFPs \cite{Felleman1991}. Not shown are the diverse compositions of IFPs; each IFP arrow represents axon bundles with a range of sizes, receptors, and neurotransmitters corresponding to diverse speeds and accuracies. \textit{(Bottom)} Controller with sensor delays and lots of IFPs, described in Sections \ref{sec:DESS} and \ref{sec:IFP}. Delayed signaling is a central feature of the brain \cite{Sterling2015}, and the presence of delays in this controller motivates the presence of IFPs.}
\label{fig:intro_figure_neuro}
\vspace{-1.5em}
\end{figure}

Neural architectures have evolved to support exceptionally efficient and robust sensorimotor control, despite the fact that biological components communicate and compute more slowly and more noisily than engineered components. These differences and known neuroanatomy suggest that neural control architectures are unlike those currently found in engineering theory. Better understanding of neural architectures is beneficial for both neuroscientists and engineers, who wish to build systems that are as scalable, efficient, and robust as organisms, and with similarly constrained hardware. 

A prominent feature of neural architectures that has eluded understanding is the presence of many \textit{internal feedback pathways} (IFPs), whose role in biological control remains ambiguous. IFPs are prevalent in neuronal systems \cite{Felleman1991, Callaway2004, Muckli2013, Suga2008} and can also be found in the immune system \cite{Busse2010,Lund2008,Palm2007} and many others; for a more detailed survey of IFPs in biology, refer to our companion paper \cite{Paper1}. Survival-critical behaviors, such as reflexive escape and object-tracking, are thought to be generated by fast forward pathways, shown in black in the top panel of Fig. \ref{fig:intro_figure_neuro} for primate visual cortex. However, physiological evidence tells us that the IFPs are more complex and diverse in speed and composition than forward paths, far more so than can be shown in a single figure. Such massive and diverse IFPs are a mystery since they have no obvious function for fast reflexes. Nonetheless, massive and diverse IFPs are found in almost all nervous systems across species and life stages. \textbf{We require new theory that explains the presence, complexity, and function of IFPs}.

Another feature of neural architectures that remains somewhat mysterious is the vast diversity in morphology, physiology, and biochemistry of the underlying components. Due to energy constraints, these components often face \textit{speed-accuracy tradeoffs} (SATs); they are either fast and inaccurate or slow and accurate but not both \cite{Sterling2015}. Recent work suggests that diverse hardware-level components (which obey SATs) and layered architectures enable performance that overcomes the hardware-level SATs \cite{Nakahira2021}. For instance, organism behavior is both fast and accurate despite cells and neurons obeying SATs. We call this a \textit{Diversity-Enabled Sweet Spot} (DESS). 

\begin{figure}
\centering
\includegraphics[width=7.5cm]{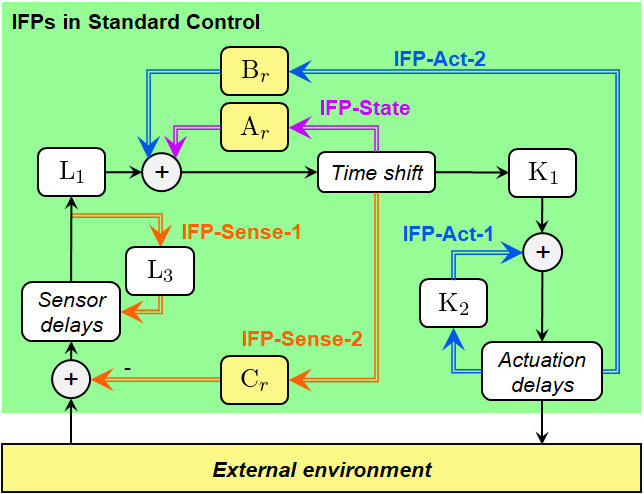}
\caption{IFPs in an output feedback controller with sensing and actuation delays. $A$, $B$, and $C$ represent the state, actuation, and sensing matrices of the physical plant; $K_1, K_2, L_1, L_3$ are submatrices of the optimal controller and observer gains. Delayed sensing and actuation induce two independent IFPs: IFP-Sense-1, and IFP-Act-1, which can be studied separately in the full control and state feedback subproblems. The remaining three IFPs (IFP-Sense-2, IFP-State, IFP-Act-2) are intrinsic to the output feedback controller.}
\label{fig:of_ifp}
\vspace{-1.5em}
\end{figure}

\textbf{We claim that IFPs are crucial architectural features that strategically combine highly diverse components to give rise to optimal performance}; that they enable DESS. In this paper, we demonstrate this using a simple but striking model using slight modifications to existing control theory. We show that the presence of delayed sensing and/or actuation gives rise to new sources of IFPs in a standard output feedback controller, as shown by ``IFP-Sense-1'' and ``IFP-Act-1'' in Fig. \ref{fig:of_ifp}. These IFPs are crucial for optimal performance, and can be studied separately in a full control problem with delayed sensing, or state feedback problem with delayed actuation. We set up a full control problem in Section \ref{sec:Methodology}; combining accurate-but-delayed sensors and fast-but-inaccurate sensors. We then demonstrate the presence and importance of DESS (Section \ref{sec:DESS}) and IFPs (Section \ref{sec:IFP}), and discuss the implications for general standard controllers. Concluding remarks are given in Section \ref{sec:Discussions_and_Conclusions}.

This paper focuses largely on the concepts of DESS and delay in requiring IFPs. We capture key elements of the problem using tools from standard control. We extend this work in our companion paper \cite{Paper3}, where we utilize System Level Synthesis (SLS) \cite{Anderson2019} theory to show that incorporating additional features of biological systems (i.e. local and delayed communications) gives rise to optimal controllers containing a much greater complexity of IFPs. Unlike \cite{Nakahira2019}, we do not include new experimental data, but we aim to provide theory that assists in guiding and interpreting existing \cite{Musall2019} and future experiments; we discuss potential experiments and biological systems of interest in our companion paper \cite{Paper1}.

%% file: tex_files/04_methodology.tex
\section{Case Study} \label{sec:Methodology}
We now present the setup of a system with diverse sensing, where we will analyze DESS (Section \ref{sec:DESS}) and IFP (Section \ref{sec:IFP}). The goal of this analysis is not to model any biological system in detail, but to study simple examples that give rise to fundamental theory on the existence and characteristics of DESS and IFP using standard control theory.

We consider a system of $n$ nodes in a ring formation. 
State dynamics in this ring can be described by the difference equation $x_r(t+1) = A_rx_r(t)$, where $x_r \in \mathbb{R}^n$ represents ring states and $A_r \in \mathbb{R}^{n \times n}$ is:

\begin{equation}
    A_r = \frac{a}{3} *
    \begin{bmatrix} 
    1 & 1 & 0 & \dots & 1\\
    1 & 1 &  1 & 0 &  \dots  \\
    0  & \ddots & \ddots &  \ddots &  \vdots \\
    \vdots & \ddots & \ddots & \ddots & 1\\
    1 &     0   & \dots & 1 &  1
    \end{bmatrix}
\end{equation}

\noindent Here $a$ determines the spectral radius of $A_r$; $a=1$ leads to a neutrally stable system, and $a>1$ to an unstable system.


We consider two types of sensors: fast and sparse, and slow and dense. This is reminiscent of biological components which obey speed-accuracy tradeoffs (e.g. proprioception vs. vision, retinal fovea vs. surround). These sensors lie on opposite ends of the speed-accuracy spectrum. We focus on the case of perfect actuation and diverse sensing, i.e. full control (FC), and assume zero sensor noise for simplicity\footnote{We assume zero sensor noise purely for pedagogical purposes; our findings on DESS and IFP are replicable with sensor noise, although the associated plots are slightly less cleanly interpretable.}. All analysis and findings naturally apply to the dual case of perfect sensing and diverse actuation, i.e. state feedback (SF). We will also briefly discuss the case of imperfect actuation and sensing in output feedback (OF) at the end of Section \ref{sec:IFP}, and a simple extension to standard separation theory.

Consider the discrete-time linear time-invariant system:
\begin{equation} \label{eq:FC}
\begin{aligned}
x(t+1) & = Ax(t) + B_2u(t) + B_1w(t) \\
y(t) & = Cx(t)
\end{aligned}
\end{equation}
\noindent where $x$, $u$, $w$, and $y$ are the state, control action, disturbance, and output, respectively.

To model internal sensing delays, we add delay states to the model. Assume we sense each state, but with a delay of $d$ timesteps; then, we introduce $d$ copies of each ring state to represent information internal to the delayed sensors. The resulting state vector is $x = \begin{bmatrix} x_r^\top & x_s^\top \end{bmatrix}^\top \in \mathbb{R}^{n(d+1)}$. $x_r$ corresponds to the original ring states; $x_s$ represent delayed sensor states. Note: we define $x_s$ as being internal to the agent; these correspond to the arrows exiting the delay blocks in Fig. \ref{fig:intro_figure_neuro}. We can think of $x_s$ as information sensed by the eye that is being passed along a delayed visual pathway to LGN, then V1, and so on. The motor areas of the brain can only access the most delayed information, i.e. the values of $x_s$ corresponding to a delay of $d$; these will be picked out by the `sensing' matrix $C$.

For full control, we define the input vector $u = \begin{bmatrix} u_r^\top & u_s^\top \end{bmatrix}^\top \in \mathbb{R}^{n(d+1)}$; this has dimension equal to $x$. $u_r$ represents physical actuation upon the ring states, and $u_s$ denotes internal wires to internal delayed states; this corresponds to the red IFP arrows exiting the $L$ blocks in Fig. \ref{fig:intro_figure_neuro}. Although both $u_r$ and $u_s$ are control signals in the standard sense, they have vastly different interpretation and engineering cost; $u_r$ represents physical actuation, which for an organism requires high-cost muscle cells, etc, while $u_s$ represents low-cost internal communication, i.e. neurons.

The state matrix $A$ for the vector $x$ including delay states has dimension $n(d+1) \times n(d+1)$, and can be written as 
\begin{equation}
    A = \begin{bmatrix}
    \hat{A}_r & 0_{n \times n} \\
    I_{nd \times nd} & 0_{nd \times n}
    \end{bmatrix}
\end{equation} 
\noindent where $\hat{A}_r = \begin{bmatrix} A_r &  0_{n \times n(d-1)}\end{bmatrix}$ has dimension $n \times nd$, and $I$ and $0$ are identity and zero matrices. $A$ has equal spectral radius as $A_r$; its stability is also determined by $a$. For full control (FC), we have full actuation, i.e. $B_2=I_{n(d+1) \times n(d+1)}$. Additionally, we only consider disturbances that affect the ring states, i.e. $B_1=\begin{bmatrix} I_{n \times n} & 0_{n \times nd} \end{bmatrix}^\top$.

For fast and sparse sensing, instead of sensing each state, we choose to sense along eigenvectors of the ring state matrix $A_r$. We will choose some $q$ eigenvectors, typically corresponding to the $q$ highest-magnitude eigenvalues of $A_r$. $C_f$ has dimension $q \times n(d+1)$, and is written as $C_f = \begin{bmatrix} E & 0_{q \times nd} \end{bmatrix}$, where the rows of $E \in \mathbb{R}^{q \times n}$ are the $q$ selected eigenvectors of $A_r$. $C_f$ represents instant sensing which senses along the $q$ ``worst directions'' of impulse propagation. For slow and dense sensing, we sense each state with a delay of $d$ timesteps. $C_s$ is of size $n \times n(d+1)$, and can be written as $C_s=\begin{bmatrix} 0_{n \times nd} & I_{n \times n} \end{bmatrix}$. We will study the system with fast-only, slow-only, and diverse ( both fast and slow) sensors, corresponding to $C=C_f$, $C=C_s$, and $C=\begin{bmatrix} C_f^\top & C_s^\top \end{bmatrix}^\top$, respectively.

We consider a standard LQR cost with state penalty $Q=B_1 B_1^\top$. Note that we only penalize the $n$ ring states, not the delayed sensing states. The optimal controller is $u(t) = -Ly(t)$, where $L = A^\top SC(C^\top SC)^{-1}$ for matrix $S$ that solves the DARE. Assuming no sensor noise, the LQR cost can be calculated via $\text{Tr}(B_1^\top S B_1)$ \footnote{In plots, we will normalize the LQR cost by the number of ring nodes, i.e. cost $=\text{Tr}(B_1^\top S B_1)/n$.}. We now use this system to demonstrate substantial presence of both DESS and IFPs.

%% file: tex_files/05_dess.tex
\section{Diversity-Enabled Sweet Spot (DESS)} \label{sec:DESS}
As described above, we consider two sensor types: fast and sparse, and slow and dense. We first study system performance with uniform sensors; this gives terrible performance and incurs extreme costs. We then use diverse sensors, which give a dramatic performance improvement and demonstrate the existence of a DESS. For the remainder of this paper, unless otherwise specified, we use parameters $n=5$, $a=1.856$ (to be explained), $q=1$, and $d=3$.

\subsection{Fast-and-sparse sensing}
\begin{figure}
\centering
\includegraphics[width=8.5cm]{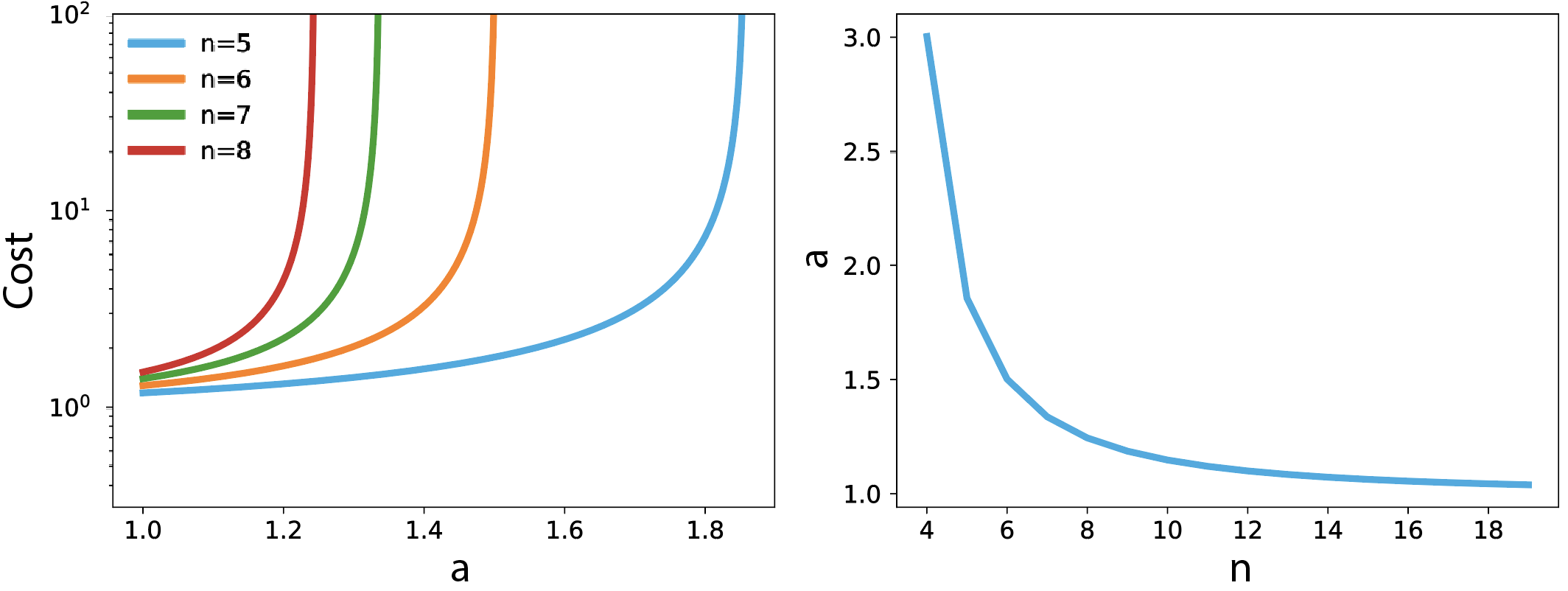}
\caption{System behavior with fast-only sensing. \textit{(Left)} Cost as a function of stability $a$ for different system sizes $n$. In all cases, the cost diverges as the system becomes less stable and the controller fails to stabilize it. \textit{(Right)} Maximum instability $a$ that can be stabilized by the optimal controller, as a function of system size $n$. As system size grows, stabilization is increasingly challenging and we are only able to deal with systems with small instability.}
\label{fig:fastonly_costs}
\end{figure}

For a controller with only fast-and-sparse sensing, we show the relationship between the cost and the stability value $a$ in the left panel of Fig. \ref{fig:fastonly_costs}. For each line, there is clearly a breaking point at which fast-only sensing results in instability and infinite cost. These breaking points occur when we reach a value of $a$ that gives more than $q=1$ unstable eigenvalues in $A_r$; in this case, fast-only sensing is no longer sufficient for the controller to reject all directions of unstable impulse propagation. Thus, certain states never converge to zero, yielding infinite cost.

The right panel of Fig. \ref{fig:fastonly_costs} shows the breaking point values of $a$ for which fast-only sensing results in infinite costs (i.e. fails to stabilize). As the system size $n$ increases, the controller is less able to stabilize unstable open-loop systems. For large system sizes, the controller can only handle marginally unstable systems (i.e. $a$ very close to 1).

For $n=5$, the minimum value of $a$ that causes instability and incurs infinite cost is $a=1.856$. The corresponding impulse response is shown on the left in Fig. \ref{fig:impulse_responses_basic} which does not get attenuated to zero, as expected.

\subsection{Slow-and-dense sensing}
\begin{figure}
\centering
\includegraphics[width=8.5cm]{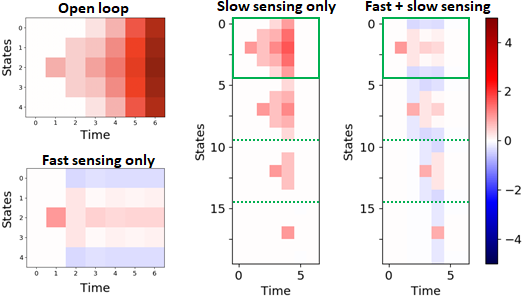}\vspace{-0.5em}
\caption{Impulse responses for the ring system. For the center and right panels, the first 5 states (outlined in solid green) correspond to physical states $x_r$, whereas later states (separated by dotted green lines) represent delayed sensor states $x_s$. \textit{(Left, Top)} Open loop, cost = $\infty$. The system is unstable. \textit{(Left, Bottom)} Fast-only, cost = $\infty$. The state does not decay to zero. \textit{(Center)} Slow-only, cost = $13.726$. The impulse response propagates until the delayed information reaches the controller, at which point all states become zero. \textit{(Right)} Diverse, cost = $2.279$. The impulse response is a mix between fast-only and slow-only cases; some portion of the impulse is rejected right away, and all states become zero after delay $d$.}
\label{fig:impulse_responses_basic}
\end{figure}

Consider a controller with only slow-and-dense sensing: the impulse response is shown in the middle panel of Fig. \ref{fig:impulse_responses_basic}. In the first three timesteps, the system behaves as if it were open-loop; no action is taken because due to delayed sensing, the controller does not yet know about the disturbance. Once the delayed information reaches the controller, the impulse propagation is immediately killed off, thanks to the dense information provided by the slow sensor.

\subsection{Diverse sensing}
Consider a controller with diverse sensors, i.e. both the fast-and-sparse and slow-and-dense sensors. The impulse response is shown in the right panel of Fig. \ref{fig:impulse_responses_basic}. 
The diverse controller displays a mix of behaviors from the fast-only and slow-only setups. In the first three time steps, the behavior is identical to the fast-only system; when information from the slow sensor reaches the controller, all impulse propagation is completed attenuated and the state values go to zero. This controller also has much-improved cost; it incurs a performance cost that is nearly six times smaller than the best non-diverse setup. This displays a dramatic DESS.

\begin{figure}
\centering
\includegraphics[width=6cm]{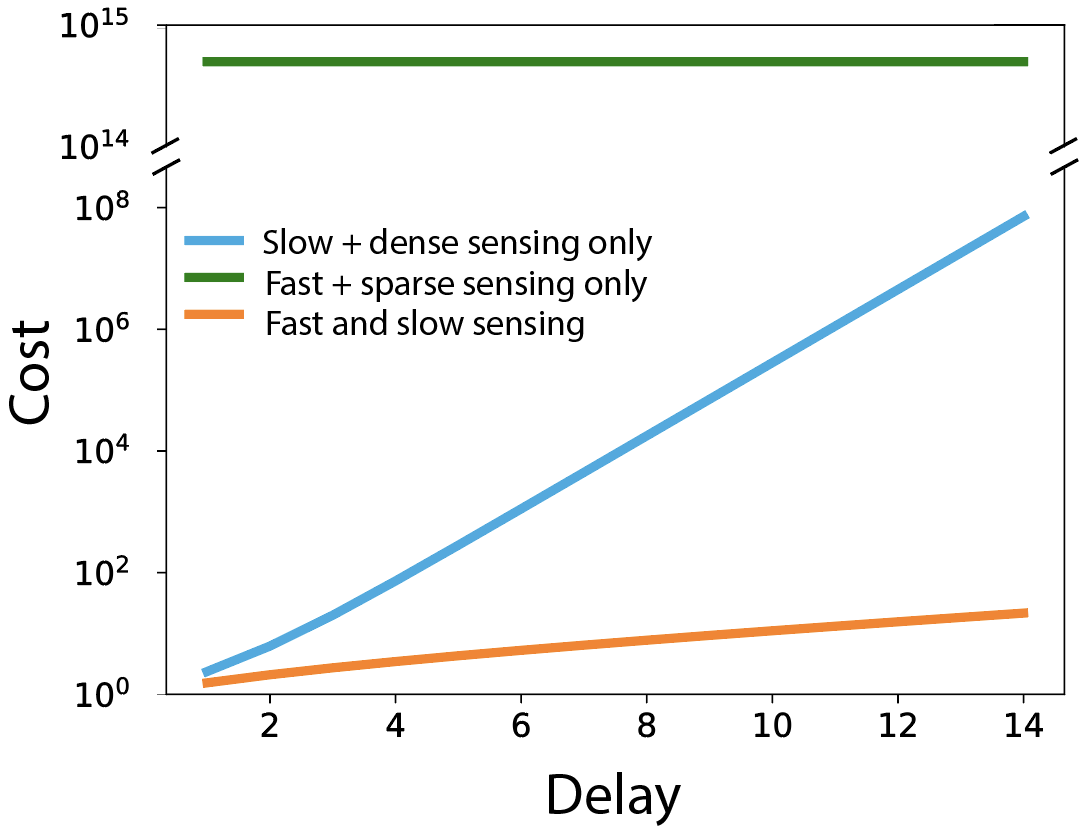}\vspace{-0.5em}
\caption{Cost of controllers using fast-only, slow-only, and diverse (i.e. fast and slow) sensing for varying values of delay. The fast-only setup incurs exorbitant cost across all values of delay. The slow-only setup performs decently for low delay, but incurs exponentially increasing costs for increasing delay. The diverse setup maintains a low cost for all delays.}
\label{fig:cost_vs_delay}
\vspace{-1em}
\end{figure}

We vary the delay parameter and show that this DESS only becomes more dramatic for increased values of delay; this is shown in in Fig. \ref{fig:cost_vs_delay}. Clearly, the diversity in sensors achieves a sweet-spot; while fast-only sensing incurs effectively infinite costs and slow-only sensing incurs dramatic exponentially increasing costs, diverse sensing allows us to keep the cost below $10^1$. Thus, delayed sensing is acceptable if paired with a fast-but-sparse sensor; similarly, sparse sensing is acceptable if paired with dense-but-slow sensor. This sweet spot is consistent with what we expect from the sensorimotor system and neural systems in general; neurons and cells typically obey SATs, but a combination of diverse neurons (i.e. both fast and slow) allows the system to achieve a DESS.

Overall, this simple but striking example suggests that DESS is a fundamental feature in the presence of diverse components. Even with only two types of sensors, we observe a strong DESS. We expect that in systems with more diversity between components (i.e. biological and neural systems), even more extreme DESS may be found.

We remark that for systems with larger values of $n$ and $a$, even more pronounced DESS is observed. This is because fast-only sensing becomes more sensitive to instability as $n$ increases, as suggested in Fig. \ref{fig:fastonly_costs}. On the other hand, increasing $q$ improves the performance of the fast-only system. As long as $q$ is less than the number of unstable eigenvalues in $A_r$, we observe dramatic DESS. When $q$ is greater or equal to the number of unstable eigenvalues in $A_r$, we still observe a DESS, albeit less drastic one. Also, all analysis and results in this section apply to the dual case of diverse actuation, in which dramatic DESS is also observed.

Considering the terrible performance of the non-diverse setups, we conclude that the diverse setup of sensors performs far better than the sum of its individual parts. This illustrates a basic principle at work: by cleverly using architecture, we can combine individual components that together perform better than the sum of their parts (i.e. achieve a DESS). We now study a fundamental architectural feature that enables this DESS -- internal feedback pathways (IFPs).

%% file: tex_files/06_ifp.tex
\section{Internal Feedback Pathways (IFPs)} \label{sec:IFP}

We now observe the structure of the controller that gives the dramatic DESS found above. We show that IFPs are an essential architectural feature that enable DESS and that removing IFPs destroys system performance.

\begin{figure}
\centering
\includegraphics[width=6.5cm]{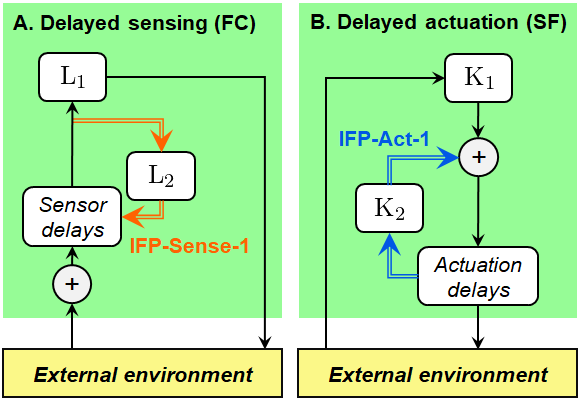}
\caption{IFPs in standard controllers with delay. Forward pathways are indicated by black arrows. \textit{(A)} Full control with delayed sensing. The IFP is shown in orange. The optimal FC controller is $L = \begin{bmatrix}L_1^\top & L_2^\top \end{bmatrix}^\top$; $L_1$ represents forward processing, while $L_2$ represents an IFP. We can further partition $L_2$ into $L_2 = \begin{bmatrix}L_{2,1}^\top & L_{2,2}^\top & L_{2,3}^\top \end{bmatrix}^\top$, as shown in the bottom panel of Fig. \ref{fig:intro_figure_neuro}. \textit{(B)} State feedback with delayed actuation. The IFP is shown in blue. The optimal SF controller is $K = \begin{bmatrix}K_1 & K_2 \end{bmatrix}$; $K_1$ represents forward processing, while $K_2$ represents an IFP.}
\label{fig:fc_ifp}
\vspace{-1em}
\end{figure}

The IFPs in controllers are defined as information from the actuation output to the sensor input. We show the presence of IFPs in our FC controller with delayed sensing and its dual (SF controller with delayed actuation) in Fig. \ref{fig:fc_ifp}. Our definition of the delay states as \textit{internal, sensor states} means that all control action to them (i.e. the orange arrow exiting from $L_2$ in Fig. \ref{fig:fc_ifp}) constitute IFPs. This corresponds to internal wires in the controller rather than actuation upon the external environment. Conversely, in a system with no internal states, a static gain controller $u=Kx$ or $u=Ly$ would contain only forward paths from sensor input ($x$ for SF and $y$ for FC) to actuation output $u$, with no IFPs.

\begin{figure}
\centering
\includegraphics[width=5.5cm]{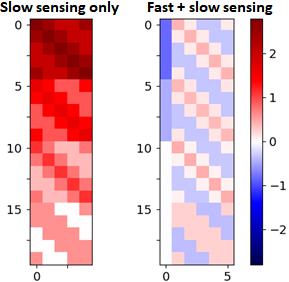}
\caption{Optimal controller $L$ for slow sensing (left) and diverse sensing (right). For both, the first 5 rows correspond to forward paths, while subsequent rows represent IFPs. In both controllers, IFPs are abundant.}
\label{fig:controller_pics}
\end{figure}

In our model, IFPs only appear in the model when the slow sensor (i.e. delayed internal states) is present, whether on its own or in a diverse-sensor setup. For the optimal FC controller $L = \begin{bmatrix} L_1^\top & L_2^\top \end{bmatrix}^\top$
\noindent where $L_1 \in \mathbb{R}^{n \times n}$, entries in $L_2$ corresponds to IFP. We show our slow-only and diverse controllers from the previous section in Fig. \ref{fig:controller_pics}. For the slow-only controller, the shape of the $A_r$ matrix is clearly visible within both the forward and IFP blocks. This is not a coincidence, since both forward and feedback (i.e. IFP) paths utilize $A_r$ to predict and counteract the behavior of the ring states. For the controller with both fast and slow sensors, the shape of the blocks are still similar, although including the fast sensor changes the magnitudes of the entries in the controller corresponding to the slow and dense sensor. 

In both controllers, the system uses IFPs to produce systems with optimal performance -- in the case of the diverse controller, IFPs are required to give the DESS discussed in the previous section. To further study the impact of IFPs, we remove the IFPs in setups with slow and diverse sensing and observe the resulting performance. This is done by setting $L_2$ to be zero in the optimal controller. We remark that this directly corresponds to experimental conditions in which an organism's IFPs are knocked out via surgery or medication. 

\subsection{System performance without IFPs}
\begin{figure}
\centering
\includegraphics[width=6cm]{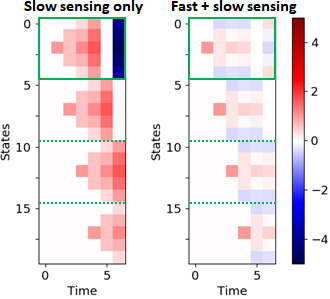}
\caption{Impulse responses of controllers with removed IFPs. Prior to removing IFPs, both controllers stabilized the system. After removing IFPs, neither stabilizes the system; cost = $\infty$ for both.}
\label{fig:no_ifp}
\vspace{-1em}
\end{figure}

The removal of IFPs severely degrades performance, as shown in Fig. \ref{fig:no_ifp}. \textit{With} IFPs, both the slow-only and diverse controller delivered stable performance. After removing IFPs, both controllers become unstable.

Removal of IFPs causes a gap in system information about previously implemented actions and current state. The resulting system acts based on what has been sensed $d$ timesteps ago \textit{without taking previous, more recent actions into account.} This is apparent in the impulse response of the slow-only setup on the left in Fig. \ref{fig:no_ifp}. The impulse is attenuated perfectly after $d+1$ time steps. However, immediately in the next timestep, more attenuating control action is received, causing the states to become negative (due to the \textit{positive} sensed state $d$ timesteps ago). The system will fluctuate between positive and negative states, each time growing in magnitude. Overall, the removal of IFPs causes the system with slow-only sensing to become unstable. 

Removing IFPs in the diverse sensing case yields similarly bad results; the impulse response again fluctuates such that the amplitude increases to infinity. Compared to the slow-only setup, the initial state magnitudes are smaller; however, states eventually diverge due to instability.

An interesting observation: if we reduce the instability value $a$ of the system, then a setup with fast-only sensing is sufficient to stabilize the system. If we add slow sensing to this to create a diverse sensing setup, then remove IFPs, the resulting system still fluctuates but eventually decays to zero instead of blowing up. In this case, the diverse setup is \textit{worse} than the fast-only setup. This again emphasizes the importance of IFPs, not only for performance but also to facilitate DESS in systems where delays are present.

\subsection{IFPs in Standard Controllers}
In the above sections, we used standard full control (FC) models to describe a simplified model with sensing delays. These results also apply to the dual state feedback (SF) case with actuation delays; both produce IFP connections, as shown in Fig. \ref{fig:fc_ifp}. An additional model to consider is output feedback (OF) where we consider a system described by

\begin{equation}
\begin{aligned}
    x_r(t+1) & = A_rx_r(t) + B_{2r}u_r(t) + B_{1r}w(t) \\
    y_r(t) & = C_rx_r(t)
\end{aligned}
\end{equation}

\noindent where $x_r \in \mathbb{R}^n$, $u_r \in \mathbb{R}^m$, and $y_r \in \mathbb{R}^p$ correspond to states, control action, and sensed outputs, respectively. The OF controller for this system inherently contains IFPs irrespective of delays being present; these are represented by `IFP-Sense-2'', ``IFP-State'', and ``IFP-Act-2'' in Fig. \ref{fig:of_ifp}. We shall refer to these as OF-IFPs. IFP-State estimates state evolution in the absence of noise and actuation, IFP-Act-2 accounts for controller action, and IFP-Sense-2 predicts incoming sensory information based on the internal estimated state. We remark that in principle, if only Bayesian estimation of the initial condition is required, no IFPs are required. However, if timely state estimations are required for action (e.g. in the Kalman filter), then OF-style IFP is necessary; thus, IFPs in OF are somewhat Bayesian in nature, they represent more than just pure Bayesian estimation. Additionally, in the adversarial $\mathcal{H}_{\infty}$ setting, an additional IFP path is necessary to anticipate the worst-case adversarial disturbance. 

As with the FC and SF cases, we can introduce internal delay states. Let the augmented state be $x = \begin{bmatrix} x_r^\top & x_a^\top & x_s^\top \end{bmatrix}^\top$, where $x_a$ represents delayed actuation states and $x_s$ represents delayed sensing states. Then, $x_a$ has length $d_am$ where $d_a$ is the number of actuation delay steps; the last $m$ elements of $x_a$ represent delayed physical actuation upon the real physical state $x_r$. Similarly, $x_s$ has length $d_sp$ where $d_s$ is the number of sensing delay steps; the last $p$ elements of $x_s$ represent delayed information from the sensor. For ease of notation, we assume that $d_a = d_s = d$; our setup trivially extends to the case that they are unequal.

We also augment the input, i.e. $u = \begin{bmatrix} u_r^\top & u_s^\top \end{bmatrix}^\top$. Now, $u_r$ represents control input which is will be delayed before reaching the physical states; $u_s \in \mathbb{R}^{pd}$ represent the internal wires to internal delay states $x_s$ (as in the FC case). 

Let $Z^{x,y} \in \mathbb{R}^{xy \times xy}$ be the block-downshift matrix, with identity matrices of size $y$ along its first block sub-diagonal and zeros elsewhere. The system matrices for the augmented system are written in block matrix form:

\begin{equation}
\begin{aligned}
    A & = \begin{bmatrix} 
        A_r & \hat{B}_{2r} & 0 \\
        0 & Z^{d,m} & 0 \\
        \hat{C}_r & 0 & Z^{d,p}
    \end{bmatrix},
    B_2 = \begin{bmatrix} 0 & 0 \\ \hat{I}_B & 0 \\ 0 & I
    \end{bmatrix}, 
    B_1 = \begin{bmatrix} B_{1r} \\ 0 \\ 0 \end{bmatrix} \\
    C & = \begin{bmatrix} 0 & 0 & \hat{I}_C\end{bmatrix} \text{, where} \\
    \hat{B}_{2r} & = \begin{bmatrix} 0_{n \times m(d-1)} & B_{2r}\end{bmatrix}, \hat{C}_r = \begin{bmatrix} C_r^\top & 0_{n \times p(d-1)} \end{bmatrix}^\top \\
    \hat{I}_C & = \begin{bmatrix} 0_{p \times p(d-1)} & I_{p \times p} \end{bmatrix}, \hat{I}_B = \begin{bmatrix} I_{m \times m} & 0_{m \times m(d-1)} \end{bmatrix}^\top
\end{aligned}
\end{equation}

We solve DAREs to obtain the optimal estimator gain $L = \begin{bmatrix}L_1^\top & L_2^\top & L_3^\top \end{bmatrix}^\top$ and optimal controller gain $K = \begin{bmatrix}K_1 & K_2 & K_3 \end{bmatrix}$. By virtue of the block-matrix structure of our system matrices, $L_2$ and $K_3$ are generally zero. This allows us to simplify the resulting OF estimator and controller into the one shown in Fig. \ref{fig:of_ifp}. For illustrative purposes, we write the simplified equations for $d=1$ below, though our observations extend to larger values of $d$:

\begin{equation}
\begin{aligned}
\delta(t+1) & = C_rx_r(t) - C_r\hat{x}_r(t) - L_3\delta(t) \\ 
\hat{x}_r(t+1) & = A_r\hat{x}_r(t) + B_{2r}x_a(t) + L_1\delta(t) \\
u_a(t) & = K_1\hat{x}_r(t) + K_2x_a(t)
\end{aligned}
\end{equation}

Here, $\delta$ is the delayed difference between the estimated sensor output and true sensor output, discounted by the observer term $L_3\delta(t)$. The resulting controller, shown in Fig. \ref{fig:of_ifp}, contains two sources of IFPs due to delay: ``IFP-Sense-1'' and ``IFP-Act-1''. These resemble the IFPs from our FC- and SF-only delay models from Fig. \ref{fig:fc_ifp}, respectively. The remaining IFPs are intrinsic to the Kalman filter in the standard OF controller. Thus, a kind of separation principle is at play: the total controller is nearly exactly the sum of its parts (FC-delayed, SF-delayed, and OF without delay). 

We also note that the delay-motivated IFPs are much bigger in dimension than the OF-IFPs; both delay-motivated IFPs have dimension of $nd$.
Of the three OF-IFPs, IFP-State and IFP-Sense-2 have dimension $n$, while IFP-Act-1 has dimension $m$. Together, these five IFPs comprise the main sources of IFP that can be modelled using standard control; all are smaller in dimension and less complex than the SLS-based IFP described in our companion paper \cite{Paper3}.

%% file: tex_files/07_conclusions.tex
\section{Discussion} \label{sec:Discussions_and_Conclusions}

The results of this paper show that dramatic DESS in systems with diverse sensing and actuation can exist, where two poorly performing components separately create a well-functioning system when combined. The results also show the importance of IFPs within delayed components, where the system performance becomes unstable without the correct feedback information. These theoretical considerations can be applied to biological control and to other complex structures that carry out large computations. The natural next steps are to extend the IFP analysis further using SLS (as is done \cite{Paper3}), and to apply this analysis toward real biological architectures via experimentation (as proposed in \cite{Paper1}).